\documentclass[review]{elsarticle}
\usepackage{amsmath}
\usepackage{multirow}
\usepackage{lineno,hyperref}
\usepackage{subcaption}
\usepackage{graphicx}
\journal{Journal of \LaTeX\ Templates}

\begin{document}

\begin{frontmatter}

\title{Detecting Multiple Seller Collusive Shill Bidding}

%\author{Jarrod Trevathan, Claire Aitkenhead, Nazia Majadi, Wayne Read}
%\address{Griffith University, Australia}
\author[mymainaddress]{Jarrod Trevathan\corref{mycorrespondingauthor}}
\cortext[mycorrespondingauthor]{Corresponding author}
\ead{j.trevathan@griffith.edu.au}

\author[mymainaddress]{Claire Aitkenhead}

\author[mymainaddress]{Nazia Majadi}

\author[mysecondaryaddress]{Wayne Read}

\address[mymainaddress]{School of ICT, Griffith University, Australia}
\address[mysecondaryaddress]{School of Mathematical and Physical Sciences, James Cook University, Australia}
%\author{Jarrod Trevathan}
%\address{Griffith University, Australia}
%\author[mymainaddress,mysecondaryaddress]{Nazia Majadi\corref{mycorrespondingauthor}}

%\author[mysecondaryaddress]{Nazia Majadi}
%\ead{nazia.majadi@griffithuni.edu.au}


\begin{abstract}
Shill bidding occurs when fake bids are introduced into an auction on the seller's behalf in order to artificially inflate the final price. This is typically achieved by the seller having friends bid in her auctions, or the seller controls multiple fake bidder accounts that are used for the sole purpose of shill bidding. We previously proposed a reputation system referred to as the Shill Score that indicates how likely a bidder is to be engaging in price inflating behaviour with regard to a specific seller's auctions. A potential bidder can observe the other bidders' Shill Scores, and if they are high, the bidder can elect not to participate as there is some evidence that shill bidding occurs in the seller's auctions. However, if a seller is in collusion with other sellers, or controls multiple seller accounts, she can spread the risk between the various sellers and can reduce suspicion on the shill bidder. Collusive seller behaviour impacts one of the characteristics of shill bidding the Shill Score is examining, therefore collusive behaviour can reduce a bidder's Shill Score. This paper extends the Shill Score to detect shill bidding where multiple sellers are working in collusion with each other. We propose an algorithm that provides evidence of whether groups of sellers are colluding.  Based on how tight the association is between the sellers and the level of apparent shill bidding is occurring in the auctions, each participating bidder's Shill Score is adjusted appropriately to remove any advantages from seller collusion. Performance has been tested using simulated auction data and experimental results are presented.
\end{abstract}

\begin{keyword}
Shill bidding \sep online auctions \sep Shill Score \sep collusion \sep auction simulation
\end{keyword}

\end{frontmatter}

%\linenumbers

\section{Introduction}
Online auction fraud can take various forms, including (but not limited to) misrepresenting an item for sale, failing to pay for or deliver goods, selling black market items, and bid shielding \cite{8,17,22}. Shill bidding is a fraudulent activity whereby a seemingly innocent bidder (i.e., a shill bidder) uses fake bids to drive up the auction's price for the seller's benefit.  The seller can have her friends operate as shill bidders, and/or can register multiple bidder accounts for the sole intention to submit shill bids. Such behaviour disadvantages legitimate bidders as they are forced to pay more for an item in order to win the auction. Shill bidding is not permitted by commercial online auctioneers, and severe penalties can be incurred by those caught engaging in shill bidding \cite{17,29}.

In March 2001, a U.S. court charged three men for their participation in a ring of fraudulent bidding in hundreds of art auctions on eBay \cite{17}. The men created more than 40 eBay user accounts using false registration information. The fraudsters gave themselves away when it was discovered that the items were misrepresented as being of greater value. Furthermore, suspicion was raised when unrealistically high shill bids were placed. These two factors provided a firm case for prosecutors. A more recent case occurred during 2010 in the UK \cite{29}. A man used two eBay accounts. The first account was used to list a minibus for sale. He then used the second account to submit fake bids in the auction to inflate the price. The man also misrepresented the minibus by illegally reducing its mileage. He was fined \pounds 5,000 under newly introduced laws designed to combat shill bidding. These two cases highlight that shill bidders can get caught and prosecuted. However, if it were not for the misrepresentation and excessive prices, would the perpetrators have been detected?

Commercial online auctioneers claim to monitor their auctions for shill bidding activity, but are reluctant to disclose their techniques. eBay's feedback reputation system does not extend to shill bidding. The only means of recourse for a bidder that suspects she is a victim of shill bidding is to contact the auctioneer. Shill bidding detection is a relatively new area in academic literature (see [1-5, 18-20]). We proposed a solution that observes bidding patterns over a series of auctions for a particular seller, looking for typical shill bidding behavior \cite{18}. The goal is to obtain statistics regarding a bidder's conduct, and deduce a measure called a \textit{Shill Score}. The Shill Score indicates the likelihood that a bidder is engaging in shill behaviour. A bidder is given a value between 0 and 10. The closer the Shill Score is to 10, the more likely that the bidder has engaged in shill-like price inflating behaviour. The Shill Score targets core strategies that a shill bidder follows. A shill bidder that deviates too far from these characteristics is less effective, and will not significantly alter the auction outcome for the seller. The Shill Score's reputation-based approach acts as both a detection mechanism and a deterrent to shill bidders. To avoid detection, a shill must behave like a normal bidder, which in effect stops her from shilling. We also extended the Shill Score in later work to detect \textit{collusive shill bidding} (i.e., where a seller has multiple shill bidders in an auction) \cite{19,20}. Colluding shill bidders can engage in more sophisticated strategies in an attempt to reduce suspicion. 

In the first of the aforementioned real world shill bidding cases, three sellers were in collusion. In other situations a single seller might be operating under several aliases (i.e., multiple seller accounts). The purpose for taking such an approach is to reduce the suspicion on any particular seller (or shill bidder) by distributing the risk of shill bidding and making it less noticeable. Collusive seller behaviour can influence elements of the Shill Score. None of the shill detection techniques in the literature specifically target using multiple seller accounts to engage in shill bidding. In this paper, we examine data from \textit{multiple sellers} for signs of seller collusion, and determine which sellers are hosting auctions with suspicious shill behaviour (referred to as \textit{multiple seller collusive shill bidding}). The algorithm firstly identifies potential groups of colluding sellers. The degree of the association between the each member in the suspect colluding group is then determined to provide evidence regarding how suspicious the group is. A modified Shill Score is calculated for a bidder across each suspect seller's auctions. If the collective Shill Scores for the bidder across all sellers are sufficiently high, the original Shill Score is recalculated with an adjustment to remove the advantage of engaging in collusive seller behaviour.  Performance has been tested using simulated auction data and experimental results are presented.
 
This paper is organised as follows:  Section 2 discusses the online auction format and provides background on shill behaviour, the Shill Score, and ways to detect collusive shill bidding.  Section 3 defines the behaviour and strategies multiple colluding sellers can use to engage in shill bidding, and presents an algorithm to extend the Shill Score to account for colluding seller behaviour.  Section 4 shows how the proposed algorithm performs with simulated auction data.  Section 5 provides some concluding remarks and avenues for future work.

\section{Online Auction Format, Shill Bidder Strategies, and Shill Detection}

\subsection{The Online Auction Format}
The specific online auction format being investigated in this paper resembles that of an eBay auction. The auction has a predetermined start and finish time.  Bidders submit bids at any stage between the start and finish time.  Each newly submitted bid must be higher than the previously submitted bid (i.e., greater than or equal to a \emph{minimum bid increment} specified by the auctioneer). When two bids are received for the same value, the bid that arrived first is accepted. The winner is the bidder with the highest bid once the auction terminates. The winner must pay the seller the amount corresponding to the winning bid.

An auction's bid history typically contains the following information for each bid received:

\begin{center}
$<$ \textit{bid \#}, \textit{bidder id}, \textit{time}, \textit{bid amount} $>$
\end{center}

Bid \# is the number of the bid in the order of when it was received, bidder id is the identity of the bidder submitting the bid, time is the time at which the bid was received (typically down to the exact second), and bid amount is the monetary value of the bid.

Note that there are some differences in how online auctioneers display their bid histories. For example, eBay masks the bidder ids for all auctions over \$200.  A bidder’s id is replaced by two random characters from her name and padded out with *s (e.g., j*****y). There are also nuances related to automated bidding (e.g., eBay’s Proxy Bidding, uBid’s Bid Butler). However, this paper assumes that all bid history information is available (i.e., no bid masking) and that the format is standardised (i.e., no automated bidding).

\subsection{Shill Bidder Characteristics and Strategies}

We refer to the most extreme competitive shill bidding strategy as aggressive shilling \cite{18}. An aggressive shill continually outbids everyone, thereby driving up the price as much as possible. This strategy often results in the shill bidder entering many bids.

In contrast, a shill bidder might only introduce an initial bid into an auction where there have been no prior bids with the intent to stimulate bidding.  This behaviour is a common practice in traditional and online auctions. However, most people typically do not consider it fraudulent. Nevertheless it is still shill bidding, as it is an attempt to influence the price by introducing spurious bids. We refer to this as \textit{benign shilling} in the sense that the shill bidder does not continue to further inflate the price throughout the remainder of the auction \cite{18}. A benign shill bidder will typically make a ``one-off" bid at or near the beginning of the auction.

To be effective, a shill bidder must comply with a particular strategy that attempts to maximise the pay-off for the seller. We define aggressive shill bidders to have the following characteristics \cite{18}:  

\begin{enumerate}
\item \textit{Bid exclusively in auctions only held by one particular seller}. However, this alone is not sufficient to incriminate a bidder. It may be the case that the seller is the only supplier of an item the bidder is after, or that the bidder really trusts the seller (usually based on the reputation of previous dealings).  
\item \textit{High bid frequency}. An aggressive shill will continually outbid legitimate bids to inflate the final price. Bids are typically placed until the seller's expected payoff for shilling has been reached, or until the shill risks winning the auction (e.g., near the termination time or during slow bidding).  
\item \textit{Few or no winnings for the auctions participated in} (as the shill's goal is to lose).  
\item \textit{Bid within a small time period after a legitimate bid}. Generally a shill wants to give legitimate bidders as much time as possible to submit a new bid before the auction's closing time.  
\item \textit{Bid the minimum amount required to outbid a legitimate bidder}. If the shill bids an amount that is much higher than the current highest bid, it is unlikely that a legitimate bidder will submit any more bids and the shill will win the auction.  
\item \textit{Bid more near the beginning of the auction}. A shill's goal is to try and stimulate bidding, by bidding early a shill can influence the entire auction process compared to a subset of it. Furthermore, bidding towards the end of an auction is risky as the shill could accidentally win. 
\end{enumerate}

Regardless of the strategy employed (i.e., aggressive or benign), a shill will still be a bidder that often trades with a specific seller but has not won any auctions. Another factor that affects a shill bidder's strategy is the value of the current bid in relation to the reserve price (i.e., a price specified by the seller as being the minimum value for which s/he will accept for the sale to proceed). For example, once bidding has reached the reserve price, it becomes more risky to continue shilling. However, this is conditional on whether the reserve price is a realistic valuation of the item that all bidders share.

\subsection{Shill Bidding Example}
Table 1 illustrates an example auction with three bidders. Each bidder is denoted as $b_1$, $b_2$, and $b_3$ respectively. Bidders $b_1$ and $b_3$ are legitimate, whereas $b_2$ is a shill bidder. $b_2$ engages in aggressive shill behaviour by outbidding a legitimate bid by the minimal amount required to stay ahead and within a small time period of the last bid.  $b_2$'s bids force the other bidders to enter higher bids in order to win. If $b_2$ was not participating in this auction, $b_1$ would have only needed to pay \$21 in order to win. Instead, $b_2$ caused $b_1$ to pay \$33, thus the shill has inflated the price by \$12.

\begin{table}[h]
\caption{An example auction with one shill bidder inflating the price for the seller.}
\centering
\begin{tabular}{|c|c|c|c|}
\hline
\textbf{Bid \#} & \textbf{bid} & \textbf{Price} & \textbf{Time} \\ \hline
15              & $b_1$           & \$33           & 20:03         \\ \hline
14              & $b_2$ (Shill)   & \$32           & 12:44         \\ \hline
13              & $b_1$           & \$31           & 12:42         \\ \hline
12              & $b_2$ (Shill)   & \$26           & 5:05          \\ \hline
11              & $b_1$           & \$25           & 5:02          \\ \hline
10              & $b_2$ (Shill)   & \$21           & 2:47          \\ \hline
9               & $b_3$           & \$20           & 2:45          \\ \hline
8               & $b_2$ (Shill)   & \$15           & 1:07          \\ \hline
7               & $b_1$           & \$14           & 1:05          \\ \hline
6               & $b_2$ (Shill)   & \$9            & 0:47          \\ \hline
5               & $b_3$           & \$8            & 0:45          \\ \hline
4               & $b_2$ (Shill)   & \$6            & 0:20          \\ \hline
3               & $b_3$           & \$5            & 0:19          \\ \hline
2               & $b_2$ (Shill)   & \$2            & 0:06          \\ \hline
1               & $b_1$           & \$1            & 0:05          \\ \hline
\end{tabular}
\end{table}

In this example, $b_2$ exhibits the typical shill behaviour described above. This is evidenced by: (i) High frequency of bids (i.e., $b_2$ has submitted more bids than both the other bidders); (ii) Has not won the auction despite the high number of bids; (iii) Is quick to bid after a legitimate bidder; (iv) Only bids the minimal amount to stay in front (i.e., only \$1 each time); and (v) Commenced bidding early in the auction, but ceased participation well before the auction ended.

\subsection{Shill Bidding Detection}

\subsubsection{Approaches to Shill Detection and Related Work}
Commercial auctioneers claim to have a substantial methodology in place for detecting shill bidding activity. However, the commercial auctioneers do not reveal exactly what these measures are to the public. Some auctioneers may possibly look at a bidder’s IP address to see whether two or more accounts are being used to submit bids from the same computer (or the seller and a bidder(s) are using the same computer).  This would indicate that collusion is occurring.  Another approach might be to check bidders’ feedback records as accounts created for the purpose of shill bidding might have limited or no feedback.  However, there seems to be numerous problems with these approaches.  

Firstly, checking someone's IP address over time could be considered a breach of privacy.  Furthermore, a shill bidder may fake her IP address to avoid detection or frame an innocent person.  In addition, it may be the case that the computer is located in an Internet cafe or a place where the computer is shared.  One person may bid then log off, and then a subsequent person comes along and uses the same computer to bid.  Additionally, computers are dynamically assigned IP addresses on networks.  A shill bidder can get around this method of detection by disconnecting and then reconnecting his/her computer to the network. Each time the shill bidder does this, she will be assigned a different IP address.

With regard to checking feedback, there are generally two instances where auction participants could be considered suspect:

\begin{enumerate}
\item User IDs with zero feedback; and
\item A pattern of the same group of users bidding on different auctions by one seller. 
\end{enumerate}
 
However, there are legitimate users with zero feedback (i.e., newly joined bidders, or those who have not yet won or participated in any auctions). There are also legitimate reasons that the same group of users might bid on different auctions by the same seller. If a seller has a good reputation, loyal bidders may choose to continually deal with the seller. Additionally, an abundance of good feedback can be misleading as some sellers can engage in a practice referred to as \textit{reputation stacking} \cite{8,17}. Reputation stacking is where the seller creates multiple bidder accounts and holds a large number of auctions for low-valued items just for the purpose of generating positive feedback.

One of the first academic approaches to dealing with shill bidding is by Wang et al. \cite{26,27}. They suggest that listing fees (referred to as a \textit{shill proof fee}) could be used to deter \textit{reserve price shilling}. Reserve price shilling is a situation that occurs when a seller is charged a fee based on what the stated reserve price is. Therefore, a seller will list a lower reserve price to avoid higher fees and then use shill bids to push the price up. Wang et al.'s proposal charges a seller an increasing fee based on how far the winning bid is from the reserve price. The idea is to compel the seller into stating her true reserve price, thereby eliminating the economic benefits of reserve price shilling. However, this method is untested and does not apply to auctions without reserve prices.

Most previous work on shill detection in online auctions is based on analysing large volumes of historical auction data to search for shill patterns. Kauffman and Wood \cite{9} used a statistical approach to detecting shill bidding behaviours and showed how the statistical data of a market would look if opportunistic behaviours do exist. They also showed how to use an empirical model to test for questionable behaviours. However, one limitation of the approach is the need to review multiple auctions over a long period of time.  Furthermore, since the statistical approach was based on analyzing a large amount of historical auction data, it was not applicable to directly analysing a particular auction where shilling behaviours might be involved.

Chau et al \cite{30} proposed a shill bidding detection method called 2-Level Fraud Spotting, which can be used to detect fraudsters in online auctions using data mining techniques by investigating historical data from eBay auctions.

Xu and Cheng \cite{29} propose an approach to detect shill suspects in concurrent online auctions (where multiple auctions for identical items are simultaneously taking place). Their auction model can be formally verified using a model checker according to a set of behavioural properties specified in pattern-based linear temporal logic. Dong et al. \cite{31,32} extend on this work by verifying shill suspects using Dempster-Shafer theory of evidence. They use eBay auction data to validate whether using Dempster-Shafer theory to combine multiple sources of evidence of shilling behaviour can reduce the number of false positive results that would be generated from a single source of evidence. Later, Dong et al. \cite{33} study the relationship between final prices of online auctions and shill activities in eBay auctions. They train a neural network using features extracted from item descriptions, listings and other auction properties. The likelihood of shill bidding is determined by the aforementioned Dempster-Shafer shill certification technique. Goel et al. \cite{34} introduce an approach for verifying shill bidders using a multi-state Bayesian network, which supports reasoning under uncertainty. They describe how to construct the multi-state Bayesian network and present formulas for calculating the probabilities of a bidder being a shill and being a normal bidder.

Some approaches have been proposed to detect shill bidding in real-time (i.e., while an auction is in progress) \cite{35,36,37,38,39}. The motivation is that actions can be taken to penalise the seller or shill bidder before the auction terminates to ensure that innocent bidders do not become victims. Such actions can include suspending or cancelling an auction, economic penalties, and account suspension or cancellation.  However, a problem with a purely real-time shill detection method is that there is insufficient information available from just one auction. A bidder's historical behaviour must be to some extent taken into account to provide sufficient evidence of shill bidding.  The real-time proposals so are merely demonstrations of a method, but lack any sort of testing to prove their effectiveness. Furthermore, the shill behaviours outlined in these papers are arbitrary and do not share a consensus amongst the academic community about what actually constitutes shill bidding.

Beranek et al. \cite{2} introduce a trust model based on reputation (user’s evaluation after performed transactions) and on examination of properties of possible fraudulent behaviour in online auctions.  The evidence is expressed and combined using belief functions. Beranek and Knizek \cite{3} extend the fraud detection approach by using contextual information whose origin is outside online auction portals. The suggested model integrates information from auctions and relevant contextual information with the aim to evaluate the behaviour of certain sellers in an online auction and determine whether it is legal or not. However, this proposal does not focus specifically on shill bidding.

\subsubsection{The Shill Score Reputation System}
In previous work \cite{18}, we proposed a reputation system that observes a bidder's bidding patterns over a series of auctions for a particular seller. The goal is to obtain statistics regarding a bidder's conduct with the seller, and calculate a Shill Score. Potential auction participants can observe other bidders' Shill Scores to determine the likelihood that any of them are engaging in price-inflating behaviour.

The Shill Score targets core shill bidding strategies.  A shill that deviates too far from these strategies is less effective, and will not significantly alter the auction outcome. This approach acts as both a detection mechanism and a deterrent to shill bidders. To avoid detection, a shill must behave like a normal bidder, which essentially restricts her ability to engage in shill bidding.

The Shill Score basically works as follows (see \cite{18} for specific details):  A bidder $b_i$, is examined over $k$ auctions held by the same seller for the behaviour outlined in Section 2.2. Each characteristic of shill behaviour is assigned a rating, which is combined to form $b_i$'s Shill Score. The Shill Score gives $b_i$ a value between 0 and 10. The closer the Shill Score is to 10, the more likely that $b_i$ is engaging in price-inflating behaviour. The algorithm's goal is to determine which bidder(s) is most inclined to be the shill out of a group of $l$ bidders. The Shill Score behavioural ratings are determined as follows: 

\begin{itemize}
\item \textbf{$\alpha$ Rating} -– Percentage of auctions bi has participated in.
\item \textbf{$\beta$ Rating} –- Percentage of bids bi has made out of all the auctions participated in.
\item \textbf{$\gamma$ Rating} -– Normalised function based on the auctions bi has won out of the auctions participated in.
\item \textbf{$\delta$ Rating} -– Normalised inter bid time for bi out of the auctions participated in.
\item \textbf{$\epsilon$ Rating} -– Normalised inter bid increment for bi out of the auctions participated in.
\item \textbf{$\zeta$ Rating} – Normalised time bi commences bidding in an auction.
\end{itemize}

Each rating is between 0 and 1, where the higher the value, the more suspicious the bidder. A bidder's Shill Score is calculated as the weighted average of these ratings:
\begin{equation} \nonumber
score=\frac{\omega_1\alpha+\omega_2\beta+\omega_3\gamma+\omega_4\delta+\omega_5\epsilon+\omega_6\zeta}{\omega_1+\omega_2+\omega_3+\omega_4+\omega_5+\omega_6} \times 10
\end{equation}
where  $\omega_i$, 1 $\leq \omega_i \leq$ 6, is the weight associated with each rating. Tsang et al \cite{24} propose what they feel are the optimal selections for weight values.  If a bidder wins an auction, then his/her $\alpha$, $\beta$, $\delta$, $\epsilon$ and $\zeta$ ratings are 0 for the particular auction (as the shill’s goal is to lose).

\subsubsection{Shill Bidding Detection Involving Colluding Shill Bidders}

The Shill Score (as outlined in \cite{18}) considers only the basic scenario with \textbf{one seller} and \textbf{one shill bidder}. In the US case described in Section 1 \cite{17}, there were three sellers, which used 40 different aliases (40 shills in effect). The sellers understood that there was less chance that they would get caught if they used multiple bidder accounts to take alternating turns at submitting shill bids. This makes it more difficult for authorities to determine which bidders are shills, as collusive behaviour allows shill bidders to appear to be more like regular bidders. In some cases, geographical proximity can be an indication of collusion if there are several shills within a close area that participate in the auction.  For example, in the shill case \cite{17}, two of the men were from California and the other was from Colorado. However, this is not a reliable indicator of shill bidding and may raise privacy concerns, as it requires examining the registration database for such relationships.

We refer to the strategies a group of shill bidders can engage in as \textit{collusive shill bidding} \cite{19,20}. We investigated approaches to shill bidding involving one seller who controls multiple shill bidders and what effect this has on the Shill Score. The main goal of shilling is to drive up the price of an item. In the situation where there is only one shill bidder, the shill's secondary goal is to attempt to do this in such a manner that it minimises her Shill Score. When there is more than one shill bidder, there are particular strategies that the group (of shill bidders) can engage in to influence some factors contributing to their individual Shill Scores. Therefore, the group's collective goal (secondary to shilling) is to minimise each member's Shill Scores.

Despite being able to use more complicated strategies, the group as a whole must still conform to certain behaviour in order to be effective as a shill. With regard to the Shill Score, all that shill bidders can do by colluding is to reduce their $\alpha$ and $\beta$ ratings. The $\gamma$, $\delta$, $\epsilon$ and $\zeta$ ratings are still indicative of shill bidding. For example, none of the colluding shill bidders will be inclined to win an auction.  Furthermore, it is still in the group's interests to bid quickly, and by minimal amounts to influence the selling price. Therefore, inter bid times and increments will be consistent for all shill bidders. Shill bidders will also bid early in an auction and cease bidding well before the end of an auction.

\begin{table}[h]
\centering
\caption{An example auction with two colluding shill bidders alternating their bids in order to reduce suspicion.}
\begin{tabular}{|c|c|c|c|}
\hline
\textbf{Bid \#} & \textbf{bid} & \textbf{Price} & \textbf{Time} \\ \hline
15              & $b_1$           & \$35           & 20:03         \\ \hline
14              & $b_2$ (Shill)   & \$32           & 12:44         \\ \hline
13              & $b_1$           & \$31           & 12:42         \\ \hline
12              & $b_3$ (Shill)   & \$26           & 5:05          \\ \hline
11              & $b_1$           & \$25           & 5:02          \\ \hline
10              & $b_2$ (Shill)   & \$21           & 2:47          \\ \hline
9               & $b_1$           & \$20           & 2:45          \\ \hline
8               & $b_3$ (Shill)   & \$15           & 1:07          \\ \hline
7               & $b_1$           & \$14           & 1:05          \\ \hline
6               & $b_2$ (Shill)   & \$9            & 0:47          \\ \hline
5               & $b_1$           & \$8            & 0:45          \\ \hline
4               & $b_3$ (Shill)   & \$6            & 0:20          \\ \hline
3               & $b_1$           & \$5            & 0:19          \\ \hline
2               & $b_2$ (Shill)   & \$2            & 0:06          \\ \hline
1               & $b_1$           & \$1            & 0:05          \\ \hline
\end{tabular}
\end{table}

There appear to be three possible strategies that can be employed by colluding shill bidders. The first strategy is referred to as the \textit{alternating bid strategy}.  Two (or more) colluding shill bidders each take alternating turns at bidding, e.g., $shill_1$ bids, then $shill_2$ bids, then $shill_1$ bids again, etc. Table 2 presents an example of the alternating bid strategy. Here there are three bidders, denoted $b_1$, $b_2$ and $b_3$ respectively. $b_1$ is a legitimate bidder, but $b_2$ and $b_3$ are shill bidders. $b_2$ and $b_3$ take alternating turns at outbidding $b_1$. This strategy has the effect of lowering $b_2$ and $b_3$'s $\beta$ ratings (i.e., the number of individual shill bids in an auction), but does not affect their $\alpha$ ratings.  

The second strategy is for colluding shills to take turns at shilling for a particular auction (referred to as the \textit{alternating auction strategy}). For example, given two auctions, $shill_1$ will bid exclusively in $auction_1$, while $shill_2$ bids only in $auction_2$. This strategy lowers the shills' $\alpha$ ratings (i.e., number of auctions participated in), but does not affect their $\beta$ ratings.

The third strategy is to use a combination of the alternating bid and alternating auction strategies (referred to as the \textit{hybrid strategy}). The hybrid strategy can be used to alter the group's $\alpha$ and $\beta$ ratings between the two extremes. An example of a hybrid strategy would be for $shill_1$ and $shill_2$ to alternately bid in $auction_1$, $shill_3$ and $shill_4$ alternately bid in $auction_2$, then shill1 and $shill_3$ alternately bid in $auction_3$, etc. This continues until all combinations of bidders have been used, and then the process repeats. In reality, colluding shills would probably employ a hybrid strategy.

In \cite{19,20} we describe how to extend the Shill Score to detect collusive shill bidding behaviour using a \textit{Collusion Score}. While the details this approach are outside the scope of this paper, the purpose of this discussion is to highlight how differing collusive behaviours can be used in an attempt to influence the Shill Score.  

A limitation of our collusive shill bidding proposal is that it only focused on one seller who controls multiple shill bidders. There is no literature that addresses the situation where multiple sellers are in collusion and what strategies they can engage in. In this paper we refer to this scenario as multiple seller collusive shill bidding.

\section{Shill Bidding Detection Involving Multiple Seller Accounts}
This section describes the behaviours and strategies multiple colluding sellers can use to engage in shill bidding and what affect this has on the Shill Score. We then present an algorithm that takes into account multiple seller collusive shill bidding strategies and adjusts the Shill Score for a specific bidder appropriately to remove any advantage that seller collusion may have.  

\subsection{Multiple Collusive Seller Shill Bidding Behaviour and Tactics}
The goal of this paper is to address shill bidding strategies that a seller could engage in if she has control of multiple seller accounts.  In order to narrow the scope of the problem, we restrict our attention to the case where there are two or more sellers in collusion, but they only control one shill bidder.  That is, we are ignoring instances where the colluding sellers control multiple shill bidding accounts (this will be the focus of future work).

Let the set of all bidders in the auction dataset be 
\begin{equation}\nonumber
B = \{b_1, b_2, \dots , b_l\}
\end{equation}

where $b_i$ denotes the $i$th bidder and $1 \leq i \leq  l$.

The set of all sellers is
\begin{equation}\nonumber
S = \{s_1, s_2, \dots , s_m\}
\end{equation}

where $s_j$ denotes the $j$th seller account and $1 \leq j \leq m$.

The set of all auctions in the dataset is

\begin{equation}\nonumber
A = \{a_1, a_2, \dots , a_n\}
\end{equation}

where $n$ ($n \geq 0$) is the total number of auctions.

The set of auctions conducted by seller $s_j$ is

\begin{equation}\nonumber
A^j = \{a_1^j, a_2^j, \dots , a_k^j\}
\end{equation}

where $A^j \subseteq A$ and $k$ ($0 \leq k \leq n$) is the total number of auctions conducted by $s_j$. Each sellers' auctions forms a partition over $A$.  That is, $A = A^1 \cup A^2 \cup \dots \cup A^j$ and $A^i \cap A^j= \{\}$, $i \neq j$, where each $A$ is pairwise disjoint.

Let us denote a seller engaging in collusive shill bidding behaviour as $S_{shill}$. $S_{shill}$ controls a series of $w$ ($0 \leq w \leq n$) seller accounts. The challenge is to determine which subset of seller accounts in $S$ are controlled by $S_{shill}$. We will denote the colluding seller accounts as the set $S^{ꞌ}$ where $S^{'} \subseteq S$.  

Consider the case where $S_{shill}$ is using two colluding seller accounts $S^{ꞌ}= \{s_1, s_2\}$, and controls a single shill bidder $b_s$ ($b_s \in B$). If $b_s$ were to participate entirely in auctions held by $s_1$, then $b_s$'s Shill Score will be high for $s_1$, but 0 for $s_2$. This is due to the $\alpha$ rating being high in $s_1$'s auctions, but 0 in $s_2$'s auctions as $b_s$ has not participated in any auctions held by $s_2$. The same holds true if the situation is reversed and $b_s$ participates in auctions held by $s_2$ and does not participate in $s_1$'s auctions (i.e., a high $\alpha$ rating for $s_2$, but a 0 $\alpha$ rating for $s_1$).

Now consider the case where $S_{shill}$ is using two colluding seller accounts $S^{ꞌ}= \{s_1, s_2\}$, and is conducting a total of four auctions (two per seller).  That is, the set of auctions held by $s_1$ is $A^1 = \{a_1^1, a_2^1\}$, and the set of auctions held by $s_2$ is $A^2 = \{a_1^2, a_2^2\}$. In order to reduce the impact of $b_s$'s $\alpha$ rating, the best strategy for $S_{shill}$ is to alternate $b_s$ evenly between the two seller's auctions.  That is, use $b_s$ in some auctions held by $s_1$ and $s_2$, but do not use $b_s$ for all auctions held by $s_1$ or $s_2$. The optimal sequences are as follows:

\begin{table}[h]
\centering
\begin{tabular}{llll}
\multicolumn{2}{l}{\textbf{Sequence 1}} & \multicolumn{2}{l}{\textbf{Sequence 2}} \\
     & $a_1^1$ -- Contains shill bidding     &      & $a_1^1$ -- No shill bidding           \\
     & $a_2^1$ -- No shill bidding           &      & $a_2^1$ -- Contains shill bidding     \\
     & $a_1^2$ -- Contains shill bidding     &      & $a_1^2$  -- No shill bidding           \\
     & $a_2^2$  -- No shill bidding           &      & $a_2^2$ -- Contains shill bidding     \\
\multicolumn{2}{l}{\textbf{Sequence 3}} & \multicolumn{2}{l}{\textbf{Sequence 4}} \\
     & $a_1^1$ -- Contains shill bidding     &      & $a_1^1$ -- No shill bidding           \\
     & $a_2^1$ -- No shill bidding           &      & $a_2^1$ -- Contains shill bidding     \\
     & $a_1^2$ -- No shill bidding           &      & $a_1^2$  -- Contains shill bidding     \\
     & $a_2^2$ -- Contains shill bidding     &      & $a_2^2$ -- No shill bidding           \\

\end{tabular}
\end{table}

Using such a sequence to evenly distribute $b_s$'s shill bidding activity out across the auctions held by $s_1$ and $s_2$ reduces $b_s$'s $\alpha$ rating for any particular seller. That is, the average number of auctions participated in per seller is lower, therefore suspicion is lower on $b_s$ for each particular seller.  We refer to this approach as the alternating seller strategy. That is, $S_{shill}$ is alternating $b_s$ evenly across each of the auctions by $s_1$ and $s_2$ in order to avoid detection and reduce suspicion.

Now let's scale up the problem to three sellers $S^{ꞌ} = \{s_1, s_2, s_3\}$, who are hosting the following respective auctions $A^1= \{a_1^1, a_2^1, a_3^1\}$, $A^2 = \{a_1^2, a_2^2, a_3^2\}$, $A^3 = \{a_1^3, a_2^3, a_3^3\}$. To evenly reduce suspicion on $b_s$ for each particular seller, Sshill can use any variation of the following sequence $a_1^1$, $a_2^2$, $a_3^3$ provided that $b_s$ only participates in an even number of auctions from each seller. The less number of auctions bs participates in from each seller, the better.  %(try to put a bound on the number of combinations)

However, the effect of the $\alpha$ rating in the Shill Score becomes more obscure if there are an uneven number of auctions conducted by each seller. Consider three sellers $s_1$, $s_2$ and $s_3$, who are hosting the following respective auctions $A^1 = \{a_1^1, a_2^1\}$, $A^2 = \{a_1^2, a_2^2, a_3^2\}$, $A^3 = \{a_1^3, a_2^3, a_3^3, a_4^3, a_5^3\}$.  Assume that $S_{shill}$ employs the following sequence of alternating auctions $a_1^1$, $a_1^2$, $a_1^3$.  $b_s$'s $\alpha$ rating will be high for $s_1$, medium for $s_2$, and low for $s_3$. This is due to the $\alpha$ rating looking at the percentage of auctions for a particular seller that $b_s$ has participated in. For each respective seller this will be $s_1 - 50\%$, $s_2 - 33.3\%$, and $s_3 - 20\%$. Logic might seem to suggest to not shill in $s_1$'s auctions, which in this case means that the Shill Score has done its job by disrupting the usual business of shill bidding. If $b_s$ is transferred from to $a_1^1$ to perhaps $a_2^3$ as $s_3$ has a greater number of auctions, then $s_3$'s $\alpha$ rating will jump up to 40\%, which is also an undesirable outcome.

It seems that the optimal approach for the alternating seller strategy is to host an even number of auctions for each $s_j \in S^{'}$, and evenly alternate bs between each seller.  The lower the number of auctions participated in, the lower the impact of the α rating.  As such it appears that the effect of the Shill Score in its current form on multiple seller collusive shill bidding is as follows:

\begin{enumerate}
\item Participate like a regular shill bidder and shill in all auctions across all seller accounts, and bs will have a high Shill Score for all sellers she has been involved with;
\item Alternately, shill in an ad hoc manner in some auctions across some/all seller accounts, and bs will have a high Shill Score for some sellers and a lower Shill Score for other sellers. But ultimately, there will be substantial evidence of shill bidding; or
\item Host an even number of auctions across all seller accounts, and alternate bs evenly across all $s_j \in S^{'}$, then evenly reduce the suspicion on $b_s$ for all sellers $b_s$ has been involved with.
\end{enumerate}

Therefore, the Shill Score has done its job for the first two cases. It is the third case that needs addressing. Clearly the Shill Score is successful in that it forces $S_{shill}$ to change her behaviour by expending more effort to open multiple accounts, reducing the number of auctions with shill bidding, and going to careful lengths to avoid detection. The main advantage $S_{shill}$ has is to influence the Shill Score's $\alpha$ rating (i.e., the percentage of seller’s auctions the bidder has participated in). The remaining Shill Score ratings (i.e., $\beta$, $\gamma$, $\delta$, $\epsilon$, $\zeta$) are unaffected by the alternating seller strategy (provided the aforementioned assumption is in place where we are only dealing with one shill bidder being used across multiple sellers). The remainder of this section discusses an algorithm that accounts for the alternating seller strategy in an attempt to remove any remaining advantage $S_{shill}$ might have by engaging in such an approach to shill bidding.
\subsection{The Seller Collusion Algorithm}
In this section we propose the \textit{Seller Collusion Algorithm}. The algorithm is designed to achieve the following goals:

\begin{enumerate}
\item Identify potential colluding groups of sellers;
\item Ascertain the likelihood that shill bidding is occurring within the colluding group of sellers; and
\item Account for the influence on the α rating for a specific bidder’s Shill Score by proportionally adjusting the α rating’s weight.
\end{enumerate}

The Seller Collusion Algorithm has six distinct steps:

\begin{enumerate}
\item Identify which sellers' auctions $b_i$ has participated in;
\item Determine which sellers have an association based on their dealings with $b_i$ and construct a \textit{Seller Association Graph};
\item Determine which sellers are most likely to be colluding based on $b_i$'s dealings with the suspect sellers and construct a \textit{Shill Bidding Association Graph};
\item For each seller in a group of suspected colluding sellers, calculate $b_i$'s \textit{Modified Shill Score} (i.e., remove the $\alpha$ rating from consideration);
\item Check the similarities between each Modified Shill Score to determine the likelihood the sellers are operating in a group, and that they are engaging in shill bidding; and
\item Based on a severity measure (i.e., the extent that it is likely that shill bidding is occurring), proportionally adjust the $\alpha$ rating and recalculate $b_i$'s Shill Score.
\end{enumerate}

The following sections outline the specifics of each step in our proposed algorithm to combat multiple seller collusive shill bidding.

\subsubsection{Step 1 -- Identify which sellers a bidder has been involved with}
Step one in the Seller Collusion Algorithm is to identify which sellers bi has been involved with, and for each identified seller, count how many of the seller's auctions bi has participated in. This refers to only whether the bidder has bid in any of the seller's auctions, not how many bids have been submitted in a particular auction.

\begin{table}[h]
\caption{Example data for possible seller collusion -– the number of times each bidder participated in each seller’s auctions.}
\centering
\begin{tabular}{|l|l|l|l|l|l|l|l|}
\hline
\multicolumn{2}{|l|}{\multirow{2}{*}{}}         & \multicolumn{6}{c|}{\textbf{Sellers}}                                             \\ \cline{3-8} 
\multicolumn{2}{|l|}{}                          & \textbf{$s_1$} & \textbf{$s_2$} & \textbf{$s_3$} & \textbf{$s_4$} & \textbf{$s_5$} & \textbf{$s_6$} \\ \hline
\multirow{3}{*}{\textbf{Bidders}} & \textbf{$b_1$} & 3           & 40          & 30          & 2           & 0           & 0           \\ \cline{2-8} 
                                  & \textbf{$b_2$} & 0           & 0           & 2           & 43          & 25          & 30          \\ \cline{2-8} 
                                  & \textbf{$b_3$} & 0           & 16          & 18          & 0           & 5           & 0           \\ \hline
\end{tabular}
\end{table}

Table 3 presents an example of test data that will be used to help describe the Seller Collusion Algorithm.  There are three bidders $B =\{b_1, b_2, b_3\}$ and six sellers $S =\{s_1, s_2, s_3, s_4, s_5, s_6\}$. $b_1$ has participated in 3 auctions by $s_1$, 40 auctions by $s_2$, 30 auctions by $s_3$, 2 auctions by $s_4$, and no auctions held by $s_5$ and $s_6$. Similarly, Table 1 outlines the dealings $b_2$ and $b_3$ have had with the sellers in $S$.

\subsubsection{Step 2 -- Determine which sellers have an association based on their dealings with a bidder}
Step two in the Seller Collusion Algorithm is to identify potential groups of suspected colluding sellers. To do this we use the \textit{Seller Association Graph} (SAG)
\begin{equation}\nonumber
SAG = (V, E)
\end{equation}

where $V$ is the set of all vertices (i.e., the sellers) and $E$ is the set of edges (i.e., an association between sellers). One Seller Association Graph $SAG_i$, is generated for each bidder bi (where 1 $\leq i \leq$ l). Each seller is a vertex $v$ in the graph. For two sellers $s_j$, $s_k$ (where $j$, $k>$ 0, $j\neq k$), an edge $e$ will exist between their vertices $v_j$ and $v_k$, if the bidder has participated in both of their auctions. Note that this is not restricted to mean concurrent participation (i.e., the auctions are running at the same time), but any historical participation.

\begin{figure}[h]
\centering
\includegraphics[width=12cm, height=4cm]{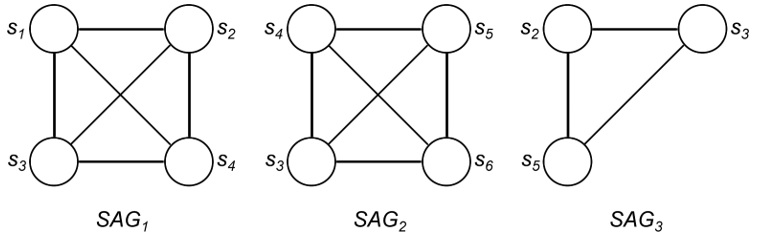}
\caption{Example Seller Association Graphs illustrating which sellers each bidder has had dealings with.}
\label{fig}
\end{figure}

Figure 1 shows the Seller Association Graphs generated based on the data in Table 3. $SAG_1$, $SAG_2$ and $SAG_3$, correspond to $b_1$, $b_2$ and $b_3$ respectively.

\begin{figure}[h]
\centering
\includegraphics[width=12cm, height=4cm]{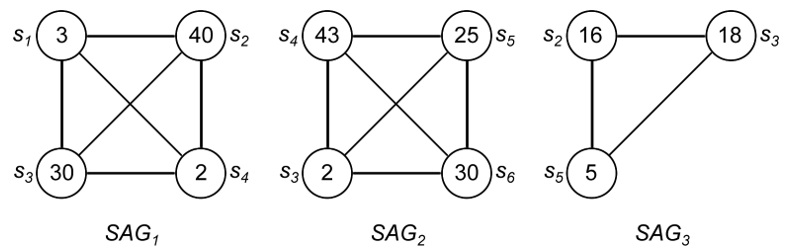}
\caption{Seller Association Graphs with weighted vertices to depict how many of each seller’s auctions a bidder has participated in.}
\label{fig1}
\end{figure}

Once each seller association graph $SAG_i$ has been defined, the respective vertices are weighted based on the number of $s_j$'s auctions bi has participated in (Figure 2). For example, in $SAG_1$, $b_1$ has participated in 3 of $s_1$'s auctions, therefore $s_1$'s vertex is weighted as 3, etc. 

\subsubsection{Step 3 -- Identify potential groups of suspected colluding sellers}

Step three in the Seller Collusion Algorithm is to determine which of the identified sellers are most likely to be in collusion with each other.  The purpose of this step is to ``weed out" more innocent sellers who might have had few (or significantly less) dealings with bi compared to more suspicious seller cliques.

In order to identify possible colluding seller groups, each edge of the Seller Association Graph is given a weighting. The weighting examines how similar (or dissimilar) each pair of sellers are in terms of the number of auctions $b_i$ has participated in with them. The first operation is to compare how far the number of auctions is from the mean number of auctions for two particular sellers $s_j$, $s_k$:

\begin{equation}\nonumber
a=(n_j - \bar{n})
\end{equation}

\begin{equation}\nonumber
b=(n_k - \bar{n})
\end{equation}

where is $\bar{n}$ is the average of the auctions held by all the sellers in $SAG_i$, and $n_j$ and $n_k$ are the values of the vertices for $s_j$ and $s_k$ respectively. Note that if $a$ or $b$ = 0 (i.e., they are equal to $\bar{n}$), then $a$ or $b$ are set to 1 to avoid division by 0 in the next equation. The following equation is used to generate the edge weighting $e_{j,k}$ for $s_j$ and $s_k$: 

\begin{equation}\nonumber
e_{j,k}=\frac{\frac{a}{|a|}+\frac{b}{|b|}-1}{\frac{a}{|a|}+\frac{b}{|b|}+1}
\end{equation}
In addition to similarities in the auctions participated in, this equation also identifies groups of sellers who have constituted a large portion of a bidder's auctions.  That is, sellers with a large number of similar participation quotas are more suspicious than similar sellers with whom bi has only participated in their auctions a small number of times. The equation will produce a value of 1 for sellers with a strong association, or -1 for sellers with a weak association.

\begin{figure}[h]
\centering
\includegraphics[width=12cm, height=4cm]{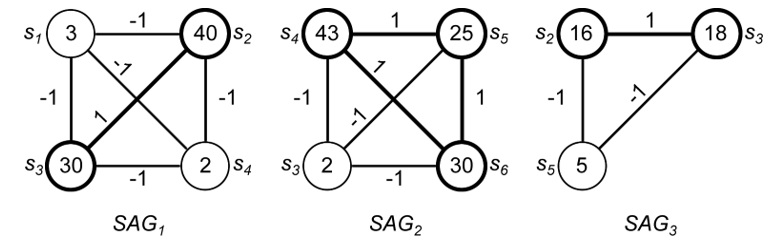}
\caption{Seller Association Graphs with weighted edges to gauge how strong the association is between groups of sellers based on the number of auctions a bidder has participated in with them.}
\label{fig2}
\end{figure}

Figure 3 shows the example Seller Association Graphs highlighting suspicious seller cliques based the edge weighting metric (i.e., those edges and vertices in bold, with an edge weighting of 1).

Those sellers which are identified as suspected colluding sellers are then added to a new graph we refer to as the Shill Bidding Association Graph (SBAG).

\begin{equation}\nonumber
SBAG = (V, E)
\end{equation}

where $V$ is a set of sellers and $E$ is a set of edges between sellers.

The corresponding Shill Bidding Association Graph for $SAG_i$ is denoted as $SBAG_i$. The non-suspect sellers' vertices and all vertex and edge weightings from $SAG_i$ are discarded in $SBAG_i$. However, $SBAG_i$ maintains all the vertices and edges for the suspect sellers that were identified in $SAG_i$.

\begin{figure}[h]
\centering
\includegraphics[width=12cm, height=4cm]{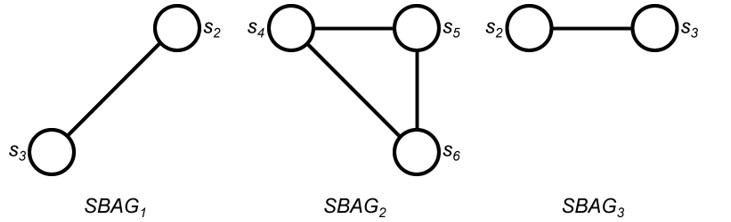}
\caption{Shill Bidding Association Graphs highlighting sellers with strong associations and with vertex weightings removed.}
\label{fig3}
\end{figure}

Figure 4 presents examples of the Shill Bidding Association Graphs. $SBAG_1$, $SBAG_2$ and $SBAG_3$, correspond to $b_1$, $b_2$, and $b_3$ respectively.

\subsubsection{Step 4 -- Calculate Modified Shill Scores to produce evidence of seller collusion}

Step four in the Seller Collusion Algorithm is to calculate $b_i$'s Shill Score for each suspected colluding seller. This will indicate which sellers are likely to have shill bidding occurring in their auctions. We would expect that a bidder who was engaging in shill bidding for a seller would have a high Shill Score. Furthermore, we would expect that the bidder would have a consistently high Shill Score across all suspect sellers in a clique.

However, given that the goal of multiple seller collusive shill bidding is to reduce the effect of the shill bidder's $\alpha$ rating, the Shill Score in its current form is unreliable. Instead, a \textit{Modified Shill Score} (MSS) is used which removes the $\alpha$ rating during the calculation of the Shill Score. That is, the MSS is calculated as though $b_i$ had only participated in $s_j$'s auctions. Each vertex in $SBAG_i$ is then weighted with $b_i$'s MSS for the particular seller.

\begin{figure}[h]
\centering
\includegraphics[width=12cm, height=4cm]{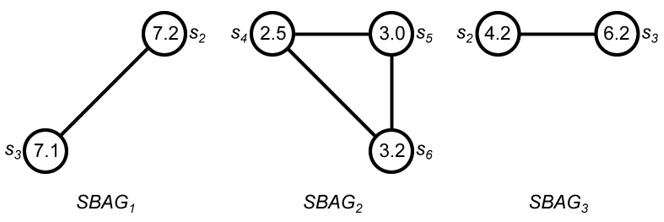}
\caption{Shill Bidding Association Graphs with Modified Shill Score weightings added to the vertices.}
\label{fig4}
\end{figure}

Figure 5 presents examples of Shill Bidding Association Graphs with the vertices weighted based on the bidder's MSS for each particular seller.  In the example, the sellers in $SBAG_1$ appear highly suspicious.

\subsubsection{Step 5 -- Classify the severity of suspected colluding sellers}

Step five in the Seller Collusion Algorithm is to classify a suspected colluding group of sellers based on the severity of shill bidding that has occurred during their auctions (i.e., those auctions for which the suspected bidder has been involved). A shill bidder is likely to have the same bidding pattern across all auctions in which they are participating. Therefore, we would expect the MSS to be similar across all auctions for the shill bidder. Sellers for which $b_i$ has a high MSS can be deemed suspect. Furthermore, sellers for which a bidder has a very similar MSS can also be considered suspicious. Sellers that fall into this category warrant further investigation.

During step four, we compare the $b_i$'s MSS for each seller. Firstly, we can remove any $s_j$ from $SBAG_i$ if $b_i$'s MSS is below 4 for $s_j$'s auctions as there is little evidence that shill bidding is occurring despite being identified as being part of a clique.

\begin{figure}[h]
\centering
\includegraphics[width=12cm, height=4cm]{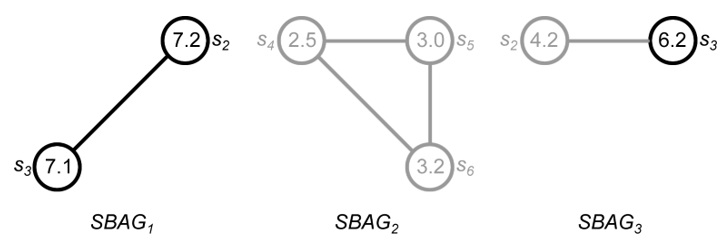}
\caption{Shill Bidding Association Graphs with sellers removed who have a Modified Shill Score below 5.}
\label{fig5}
\end{figure}

Figure 6 shows that $SBAG_2$ can now be discarded as the $b_2$ has a very low Shill Score for the auctions held by $s_4$, $s_5$ and $s_6$. $SBAG_3$ illustrates the situation where the Shill Score for $b_3$ is below the threshold of 5 for $s_2$, but is above 4 for $s_3$. In this case $s_2$'s vertex is dropped from $SBAG_3$.  

For all sellers that remain in the $SBAG_i$ we want to compare how similar their MSS scores are. To do this determine what the median MSS is for the group. The median is denoted as \emph{med}. For a given seller's vertex $v_i$ the following occurs:

\begin{center}\nonumber
If $v_i < med-0.5 \ or > med+0.5$, then discard
\end{center}

\begin{figure}[h]
\centering
\includegraphics[width=12cm, height=4cm]{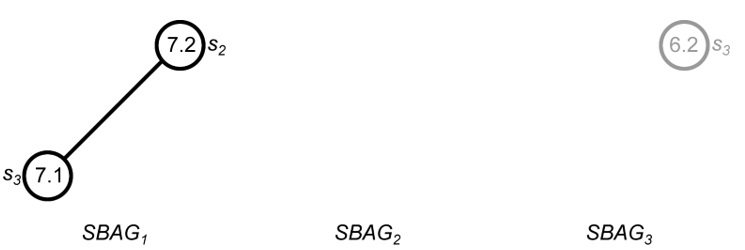}
\caption{Final Shill Bidding Association Graphs highlighting sellers that are highly suspicious.}
\label{fig6}
\end{figure}

The remaining sellers in $SBAG_i$ are now those who have sufficiently high MSSs which are relatively similar.  Therefore, these sellers have the most likelihood of engaging in the alternating seller strategy (Figure 7).  $SBAG_3$ has only one vertex remaining.  As a seller cannot be in collusion with herself, $SBAG_3$ can also be entirely discarded. However, $SGAG_1$ shows a strong association between $s_2$ and $s_3$.

\subsubsection{Step 6 -- Adjust the $\alpha$ rating of suspected shill bidders}
As previously mentioned, seller collusion is a way of avoiding shill bidding detection. The seller directly tries to influence the $\alpha$ rating of the Shill Score by increasing the number of sellers $b_i$ participates in auctions with. As such, the higher the likelihood of seller collusion, the less we can rely on $b_i$'s $\alpha$ rating. Step six in the Seller Collusion Algorithm is to adjust the $\alpha$ rating on $b_i$'s Shill Score and recalculate for the remaining vertices in $SBAG_i$.

In the original Shill Score each rating is given a weighting that influences how important the particular characteristic of shill bidding is in terms of calculating the Shill Score. The $\alpha$ rating is weighted by $\omega_1$ in the Shill Score. To mitigate against the effect of the alternating seller strategy, we need to reduce the influence of the $\alpha$ rating. This is achieved by proportionally reducing $\omega_1$ based on how high the $MSS_i$ is for $b_i$ in $s_j$'s auctions. That is, the higher $MSS_i$ the lower $\omega_1$ is for $b_i$'s $\alpha$ rating.

The first operation in achieving this outcome is to alter $b_i$'s MSS so that it is between 0 and 1:

\begin{equation}\nonumber
MSS_i = MSS_i/10
\end{equation}

Next, let $\omega^{'}_1$ denote the scaling factor to reduce $\omega_1$ by. $\omega^{'}_1$ is calculated as follows: 

\begin{equation}\nonumber
\omega^{'}_1=1-MSS_i
\end{equation}
 
Finally, the original Shill Score is recalculated for $b_i$ with the influence of the $\alpha$ rating being reduced by $\omega_1/\omega^{'}_1$:

\begin{equation} \nonumber
score=\frac{(\omega_1/\omega^{'}_1)\alpha+\omega_2\beta+\omega_3\gamma+\omega_4\delta+\omega_5\epsilon+\omega_6\zeta}{\omega_1+\omega_2+\omega_3+\omega_4+\omega_5+\omega_6} \times 10
\end{equation}

\begin{figure}[h]
\centering
\includegraphics[width=12cm, height=4cm]{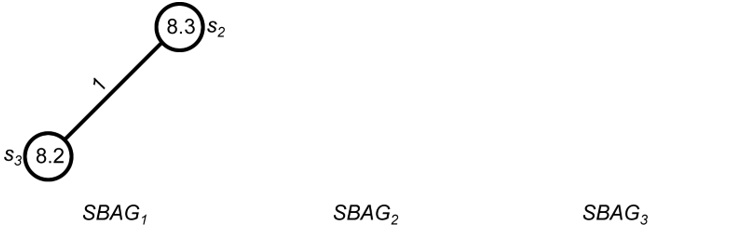}
\caption{Final Shill Bidding Association Graphs with a recalculated Shill Score adjusted for seller collusion.}
\label{fig7}
\end{figure}

Figure 8 illustrates the final results for the Seller Collusion Algorithm with the example data. $SBAG_2$ and $SBAG_3$ have been removed from consideration as there will little evidence of seller collusion. The remaining sellers in $SBAG_1$ have had $b_1$'s $\alpha$ rating adjusted to account for strong evidence of collusion.  As such $b_1$'s Shill Score with regard to these $s_2$ and $s_3$ increases as the influence of collusion on the α rating is scaled down in the Shill Score.
 
\section{Performance}
\subsection{Test Setup}
Testing the performance of any shill bidding detection technique is difficult.  As such, many shill bidding detection proposals overlook the testing stage, therefore their effectiveness cannot be ascertained.  Using data from commercial auction sites is ultimately the best test.  However, most auction sites are reluctant to supply auction data for privacy reasons and fear of lost reputation if it were discovered that shill bidding is occurring in their auctions.  Even if commercial auction data can be obtained it is still unknown if shill bidding definitely occurs as shill bidders generally are not forthcoming about their behaviour.  

The aforementioned testing problems are further exacerbated when looking for seller collusion in auction data. As this our work is the first to look into collusive behaviour, significantly larger datasets are required for testing. That is, potentially 1,000 of auctions involving numerous bidders and sellers.

Previously, we have undertaken two approaches to testing.  One is to hold simulated auctions (on a purpose-built auction server) involving human users bidding for fake items with fake money.  The users did not have any idea that there was a shill bidder in the auctions inflating the price.  However, this approach is time consuming to arrange and to continually monitor the auctions.  Additionally, such an approach may not necessarily capture all of the real-world auctioning behaviours and strategies as real money and items are not involved.

If this approach were to be used to test the Seller Collusion Algorithm, the auctions could potentially involve participants in both seller and bidder roles. During the auctions a small number of sellers would be asked to engage in seller collusion. They would not be instructed how they must do this. Once all auctions were completed the Seller Collusion Algorithm could be run against all generated auction data. However, the issues of the size of the test data set required and the amount of effort to organise the auctions really limit the viability of this approach.

The second approach to testing is to use software bidding agents to generate synthetic auction data.  Previously, we have proposed a Simple Shill Bidding Agent \cite{21} and an Adaptive Shill Bidding Agent \cite{40} that were programed to engage in typical shill bidding strategies. These agents were pitted against a set of ``Zero-Intelligence" bidding agents whose sole purpose was to randomly commence bidding up to their randomly allocated bidding limit throughout an auction.  While this approach automates the testing process and can generate large amounts of data sets, the problem was that we were programming the very behaviour we were expecting. As such, the shill detection mechanism was bound to uncover the shill bidders who were perpetrating this behaviour.

Recently, Tsang et al. \cite{23} extended upon our work and created a more sophisticated tool to generate synthetic auction data. Their approach was calibrated against data collected from the New Zealand auction site, \textit{Trade Me}, and can reproduce the statistical qualities of the auction data indicative to that of real data sets from Trade Me. 

\begin{figure}[h]
\centering
\includegraphics[width=12cm, height=6cm]{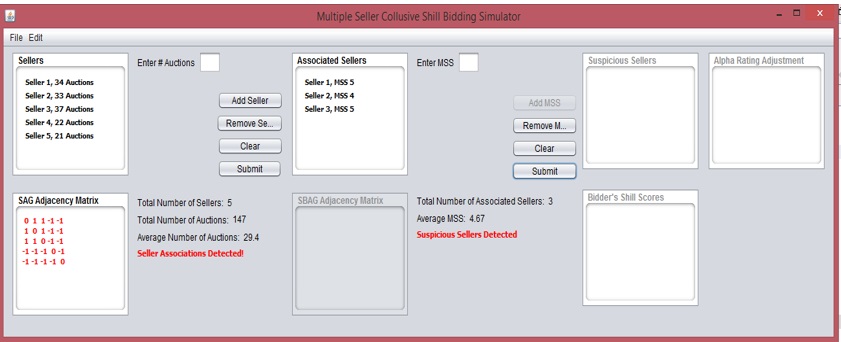}
\caption{Multiple Seller Collusive Shill Bidding Simulator.}
\label{fig8}
\end{figure}

To assist with development of the Seller Collusion Algorithm, we developed the Multiple Seller Collusive Shill Bidding Simulator. This allows a user to manually or automatically input various scenarios to automate the process of running the algorithm.  The simulator undertakes all of the calculations and provides visual output of each stage of the algorithm to ensure that the calculations are being performed accurately.

The simulator was also used as the front end for processing the synthetic auction data for the purposes of undertaking the tests.

\subsection{Results and Analysis}

The following four tests were conducted with regard to the Seller Collusion Algorithm:

\begin{enumerate}
\item How the proposed detection mechanism performs against a baseline of regular auctions that do not contain shill bidding (see Figure 10(a)).
\item How the Seller Collusion Algorithm performs on auctions involving colluding groups of sellers who engage in the alternating seller strategy to avoid detection (see Figure 10(b)).  
\item The impacts on Shill Scores and the operation of the Seller Collusion Algorithm when the shill bidder is not evenly alternated between the sellers' auctions (see Figure 10(c)).
\item Comparison between baseline Shill Scores and recalculated Shill Scores using the Seller Collusion Algorithm (see Figure 10(d)).

\end{enumerate}

\begin{figure}[h]
  \begin{subfigure}[b]{0.5\textwidth}
    \includegraphics[width=\textwidth]{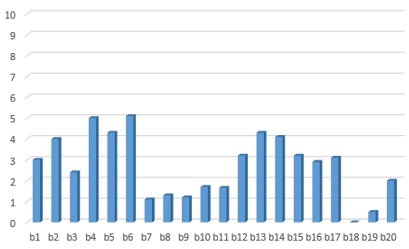}
    \caption{Shill scores in auctions without shill bidding}
    \label{fig:1}
  \end{subfigure}
  \begin{subfigure}[b]{0.5\textwidth}
    \includegraphics[width=\textwidth]{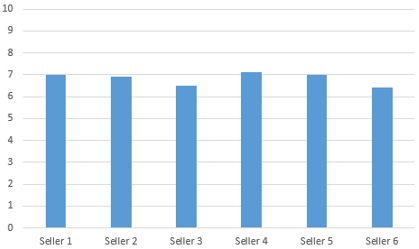}
    \caption{Shill scores using the alternating auction strategy}
    \label{fig:2}
  \end{subfigure}
  \begin{subfigure}[b]{0.5\textwidth}
    \includegraphics[width=\textwidth]{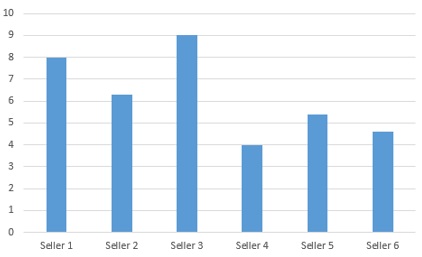}
    \caption{Shill scores not using the alternating auction strategy}
    \label{fig:2}
  \end{subfigure}
  \begin{subfigure}[b]{0.5\textwidth}
    \includegraphics[width=\textwidth]{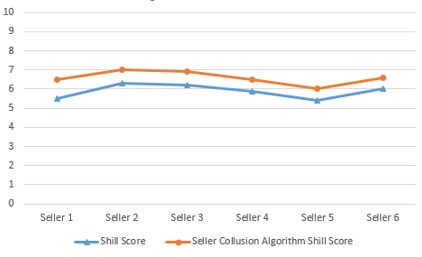}
    \caption{Baseline shill scores versus seller collusion algorithm shill score}
    \label{fig:2}
  \end{subfigure}
  \caption{Simulated Test Results.}
\end{figure}

Furthermore, we generate a simulated auction dataset using a shill bidding agent \cite{21}. The dataset contains 10 sellers, 51 bidders (including a shill bidder (e.g., `Shill\_a')), and 30 auctions. We applied the dataset on the seller collusion algorithm and compared the algorithm with collusion score approach \cite{19}. Figure 11 shows the comparative analysis results. Figure 11(a) presents that all bidders show normal bidding behaviour (including `Shill\_a'). Figure 11(b) illustrates that all bidders show regular bidding patterns except `Shill\_a' (red-filled circle). This indicates that the seller collusion algorithm performs better than collusion score approach.

We also analysed the bidding patterns of `Shill\_a' and found that `Shill\_a' achieved the highest MSSs for the sellers (e.g., `seller\_a', `seller\_b', `seller\_c', and `seller\_d') she participated in (see Figure 12). This indicates that `seller\_a', `seller\_b', `seller\_c', and `seller\_d' are the colluding sellers who engaged `Shill\_a' for price-inflating behaviour.   
\begin{figure}[h]
\centering
  \begin{subfigure}[b]{0.6\textwidth}
    \includegraphics[width=\textwidth]{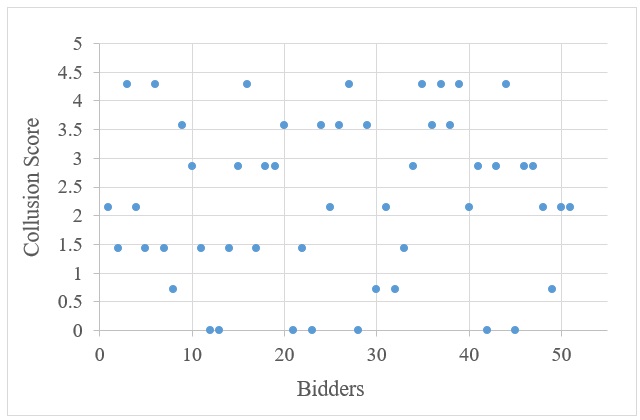}
    \caption{Collusion Score \cite{19}}
    \label{fig:1}
  \end{subfigure}
  
  \begin{subfigure}[b]{0.6\textwidth}
    \includegraphics[width=\textwidth]{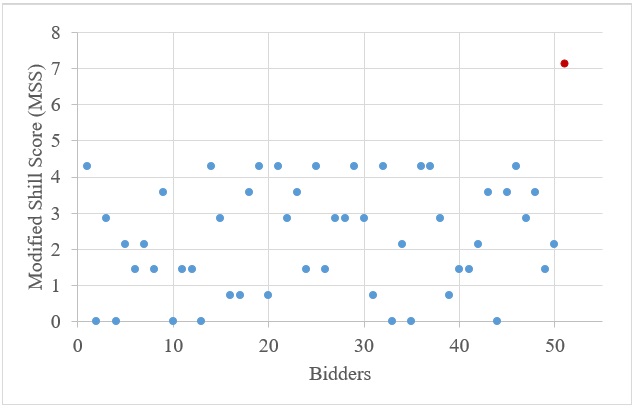}
    \caption{Seller collusion algorithm}
    \label{fig:2}
  \end{subfigure}
  \caption{Comparative Analysis Results.}
\end{figure}

\begin{figure}[h]
\centering
\includegraphics[width=10cm, height=6cm]{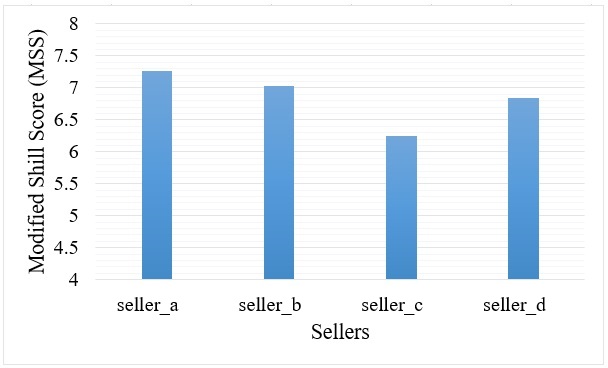}
\caption{MSSs of `Shill\_a' for the sellers she participated in.}
\label{fig8}
\end{figure}

\section{Conclusion}
This paper presented an approach to detect shilling bidding where multiple sellers are working in collusion with each other (i.e., multiple collusive seller shill bidding). In order to rein in the scope of the problem, this paper only focused on the situation where there are two or more colluding sellers who control one shill bidder.  We described the strategies colluding sellers could employ to reduce the amount of suspicion raised by shill bidding through alternating the shill bidder evenly across their auctions.  We referred to the optimal approach as the alternating seller strategy.  By engaging in the alternating seller strategy, the sellers can influence the α rating used when calculating the shill bidder’s Shill Score.

The approach to detect multiple collusive seller shill bidding outlined in this paper accomplishes two goals. Firstly, we are able to potentially ascertain colluding groups of sellers employing the alternating seller strategy.  This is achieved by identifying the sellers a bidder has been associated with (using the Seller Association Graph), determining potential cliques of sellers who might be engaging in shill bidding (using the Shill Bidding Association Graph), calculating a bidder’s Modified Shill Score for each of the sellers (i.e., removing the α rating), and ranking their likelihood of collusion.  Secondly, we are able to account for the alternating seller strategy, and recalculate the bidder’s Shill Scores for each seller.  This allows the Shill Score reputation system to remain a deterrent to those attempting to engage in shill bidding.  The information gathered by the Seller Collusion Algorithm could be supplied to auction houses, allowing them to monitor or take action against suspected sellers (i.e., warning auction participants, suspending/cancelling user accounts, suspending/cancelling auctions, economic penalties, legal action).

Performance has been tested using simulated auction data and experimental results are presented.  The tests examined how the proposed detection mechanism performs against a baseline of regular auctions that do not contain shill bidding. We then compared how our approach performed on auctions involving colluding groups of sellers who engage in the alternating seller strategy to avoid detection.  Finally, we investigated the effects of seller collusion whereby the shill bidder was not evenly alternated between the sellers’ auctions.  This algorithm has yet to be tested against live auction data.  

Future work involves investigating strategies collusive sellers can engage in whereby they control more than one shill bidder.  That is, combining the work presented in this paper with the work into collusive shill bidding \cite{19,20}.  This will significantly increase the complexity of the problem and the amount of auction data required for testing, but the goal is to create a well-rounded approach to shill bidding detection that focuses on all shill bidding strategies.  Finally, we plan to expand the Shill Score to operate in “live” auctions so that real-time actions can be taken against shill bidding perpetrators.

%\bibliography{mybibfile}

\end{document}